\documentclass[12pt]{iopart}

\usepackage{iopams}  

\newcommand{\besselj}[1]{J_{#1}}
\newcommand{\erf}{\mathrm{erf}}
\newcommand{\binom}[2]%
{\left(%
\begin{array}{c}%
      #1\\#2%
    \end{array}%
\right)}
\begin{document}

\letter%
{Fourier transform of a Bessel function multiplied by a Gaussian}

\author{Michael Carley}

\address{Department of Mechanical Engineering, University of Bath,
  Bath BA2 7AY, UK}

\ead{m.j.carley@bath.ac.uk}
\begin{abstract}
  An analytical result is given for the exact evaluation of an
  integral which arises in the analysis of acoustic radiation from
  wave packet sources:
  \begin{eqnarray*}
    I_{mn}(\beta,q) = 
    \int_{-\infty}^{\infty}
    \rme^{-\beta^{2}x^{2}-\rmi q x}x^{m+1/2}\besselj{n+1/2}(x)
    \,\rmd x,  
  \end{eqnarray*}
  where $m$ and $n$ are non-negative integers, and
  $\besselj{n+1/2}(\cdot)$ is a Bessel function of order $n+1/2$.
\end{abstract}

\submitto{\JPA}
\maketitle


In the analysis of acoustic radiation from wave packet sources in
turbulent jets~\cite{jordan-colonius13} using a series expansion of
the radiation from an asymmetric ring source~\cite{carley10}, an
integral arises which may be of more general interest, and which
appears not to be tabulated in the standard collections. It is the
Fourier transform of a Bessel function multiplied by a Gaussian:
\begin{eqnarray}
  \label{equ:integral}
  I_{mn}(\beta,q) = 
  \int_{-\infty}^{\infty}
  \rme^{-\beta^{2}x^{2}-\rmi q x}x^{m+1/2}\besselj{n+1/2}(x)
  \,\rmd x,
\end{eqnarray}
where $m$ and $n$ are non-negative integers, and
$\besselj{\nu}(\cdot)$ is a Bessel function of order $\nu$. The
integral can be evaluated exactly in terms of elementary functions
and, possibly, the error function.

From the integral definition of
$\besselj{n+1/2}$~\cite[8.411.8]{gradshteyn-ryzhik80}:
\begin{eqnarray*}
  \besselj{n+1/2}(x) =
  \frac{(x/2)^{n+1/2}}{n!\Gamma(1/2)}
  \int_{-1}^{1}
  (1-t^2)^{n}\cos xt
  \,\rmd t,
\end{eqnarray*}
and the result~\cite[3.462.3,9.253]{gradshteyn-ryzhik80}:
\begin{eqnarray*}
  \int_{-\infty}^{\infty}
  (\rmi x)^{n}
  \rme^{-\beta^{2}x^{2}-\rmi q x}\,\rmd x
  =
  -\frac{\pi^{1/2}}{2^n\beta^{n+1}}\rme^{-q^{2}/4\beta^{2}}
  H_{n}(q/2\beta),
\end{eqnarray*}
where $H_{n}(\cdot)$ is the Hermite polynomial of order $n$,
\begin{eqnarray}
\fl  I_{mn}(\beta,q) &= 
  \frac{1}{\beta(\rmi 2 \beta)^{m+n+1}}
  \frac{1}{2^{n+1/2}n!}
  \int_{-1}^{1}
  (1-t^{2})^{n}
  \rme^{-(q-t)^{2}/4\beta^{2}}
  H_{m+n+1}
  \left(
    \frac{q-t}{2\beta}
  \right)
  \,\rmd t.
\end{eqnarray}
Using the definition of the Hermite polynomial:
\begin{eqnarray*}
  H_{m+n+1}(x) = (-1)^{m+n+1}\rme^{x^{2}}
  \frac{\rmd^{m+n+1}}{\rmd x^{m+n+1}}\rme^{-x^{2}},
\end{eqnarray*}
yields, after repeated integration by parts:
\begin{eqnarray*}
  I_{mn}(\beta,q) = 
  \left.
  \frac{1}{\rmi^{m+n+1}}
  \frac{\rme^{-(q-t)^{2}/4\beta^{2}}}{\beta 2^{n+1/2}n!}
  \sum_{k=n}^{2n}
  (-1)^k 
  \frac{\rmd^{k}}{\rmd t^{k}}(1-t^{2})^{n}
  E_{n+m-k}(t)
  \right|_{-1}^{1},
\end{eqnarray*}
for $m\geq n$, where:
\begin{eqnarray*}
  \frac{\rmd^{n}}{\rmd t^{n}}\rme^{-(q-t)^{2}/4\beta^{2}} 
  = E_{n}(t)\rme^{-(q-t)^{2}/4\beta^{2}}
\end{eqnarray*}
and $E_{n}(t)$ can be evaluated using the recursion:
\numparts
\begin{eqnarray}
  E_{n}(t)
  =
  \frac{1}{2\beta^{2}}
  \left[
    (q-t)E_{n-1}(t) - (n-1)E_{n-2}(t)
  \right],\\
  E_{0}(t) = 1,\quad E_{1}(t) = \frac{q-t}{2\beta^{2}},
\end{eqnarray}
\endnumparts
which arises from Leibnitz' rule for differentiation of a product. 

The polynomial derivatives at $|t|=1$ can be found using an argument
given by Watson~\cite[page~53]{watson95}:
\numparts
\begin{eqnarray}
  \frac{\rmd^{k}}{\rmd t^{k}}(1-t^{2})^{n}
  &=
  (t)^{k-n}\frac{k!n!}{(k-n)![(2n-k)!]^{2}}2^{2n-k},
  \, n \leq k \leq 2n,\\
  &\equiv 0, \, k < n, \, k > 2n,
\end{eqnarray}
\endnumparts
so that
\begin{eqnarray}
  I_{mn}(\beta,q) &= 
  -(-2)^n
  \frac{\rme^{-(q^{2}+1)/4\beta^{2}}}{\rmi^{m+n+1}2^{1/2}\beta}\sum_{k=n}^{2n}
  \frac{1}{2^{k}}\frac{k!}{(2n-k)!(k-n)!}
  \nonumber\\
  \label{equ:result:1}
  &
  \times
  \left[
    \rme^{-q/2\beta^{2}}
    E_{n+m-k}(-1)
    -
    (-1)^k
    \rme^{q/2\beta^{2}}    
    E_{n+m-k}(1)
  \right], \, m\geq n.
\end{eqnarray}

For $m<n$, integration by parts terminates with an integral term
remaining to be evaluated:
\begin{eqnarray}
\fl  I_{mn}(\beta,q) &= 
  \frac{\rmi^{m+n+1}}{\beta 2^{n+1/2}n!}
  \int_{-1}^{1}
  \rme^{-(q-t)^{2}/4\beta^{2}}
  \frac{\rmd^{m+n+1}}{\rmd t^{m+n+1}}(1-t^{2})^{n}
  \,
  \rmd t\nonumber\\
\fl  &-(-2)^n
  \frac{\rme^{-(q^{2}+1)/4\beta^{2}}}{\rmi^{m+n+1}2^{1/2}\beta}\sum_{k=n}^{n+m}
  \frac{1}{2^{k}}\frac{k!}{(2n-k)!(k-n)!}
  \nonumber\\
\fl  &
  \times
  \left[
    \rme^{-q/2\beta^{2}}
    E_{n+m-k}(-1)
    -
    (-1)^k
    \rme^{q/2\beta^{2}}    
    E_{n+m-k}(1)
  \right].
  \label{equ:result:2}
\end{eqnarray}

Derivatives of the polynomial are given by:
\begin{eqnarray*}
  \frac{\rmd^{r}}{\rmd t^{r}}(1-t^{2})^{n}
  =
  \sum_{k=\lceil r/2 \rceil}^{n}
  (-1)^k
  \binom{n}{k}
  \frac{(2k)!}{(2k-r)!}
  t^{2k-r},
\end{eqnarray*}
with $\lceil x \rceil$ the smallest integer greater than or equal to
$x$.

Use of standard formulae~\cite[1.3.3]{prudnikov-brychkov-marichev03:1}
gives an expansion for the integral:
\begin{eqnarray}
\fl  &\int_{-1}^{1}
  \rme^{-(q-t)^{2}/4\beta^{2}}
  \frac{\rmd^{r}}{\rmd t^{r}}(1-t^{2})^{n}
  \,
  \rmd t =\nonumber\\
\fl  &
  \sum_{k=\lceil r/2 \rceil}^{n}
  (-1)^k
  \binom{n}{k}
  \frac{(2k)!}{(2k-r)!}
  q^{2k-r}
  \sum_{s=0}^{2k-r}
  \binom{2k-r}{s}
  \left(
    \frac{2\beta}{q}
  \right)^{s}
  \int_{-(q+1)/2\beta}^{(q-1)/2\beta}
  t^{s}
  \rme^{-t^2}\,\rmd t,
  \label{equ:integral:ex}
\end{eqnarray}
with $r=m+n+1$.

The integrals can be evaluated in terms of elementary functions and
the error function as:
\numparts
\begin{eqnarray}
\fl
  \int_{-(q+1)/2\beta}^{(q-1)/2\beta}
  t^{2s}
  \rme^{-t^2}\,\rmd t
  &=
  -
  \pi^{1/2}
  \frac{(2s-1)!!}{2^{s+1}}
  \left[
    \erf
    \left(
      \frac{q-1}{2\beta}
    \right)
    -
    \erf
    \left(
      \frac{q+1}{2\beta}
    \right)
  \right] \nonumber\\
\fl
  &+
  \frac{\rme^{-(q^{2}+1)/4\beta^{2}}}{2}
  \sum_{k=0}^{s-1}
  \frac{1}{2^{k}}
  \frac{(2s-1)!!}{(2s-2k-1)!!}\nonumber\\
  &\times
  \left[
    \rme^{q/2\beta^{2}}
    \left(
      \frac{q-1}{2\beta}
    \right)^{2s-2k-1}
    -
    \rme^{-q/2\beta^{2}}
    \left(
      \frac{q+1}{2\beta}
    \right)^{2s-2k-1}
  \right],\\
\fl  \int_{-(q+1)/2\beta}^{(q-1)/2\beta}
  t^{2s+1}
  \rme^{-t^2}\,\rmd t
  &=
  -\frac{\rme^{-(q^{2}+1)/4\beta^{2}}}{2}
  \sum_{k=0}^{s}\frac{s!}{(s-k)!} \nonumber\\
  &\times
  \left[
    \rme^{q/2\beta^{2}}
    \left(
      \frac{q-1}{2\beta}
    \right)^{2s-2k}
    -
    \rme^{-q/2\beta^{2}}
    \left(
      \frac{q+1}{2\beta}
    \right)^{2s-2k}
  \right].
\end{eqnarray}
\endnumparts

The integral of (\ref{equ:integral}) can thus be evaluated exactly
using (\ref{equ:result:1}) for $m\geq n$, and (\ref{equ:result:2}) for
$m < n$, with the aid of (\ref{equ:integral:ex}).

\section*{References}


\end{document}